\documentclass[doublecol]{epl2}

\title{Scalar fields in the Lense-Thirring background with a cosmic string and Hawking radiation}
\shorttitle{Scalar fields in the Lense-Thirring background with a cosmic string and Hawking radiation} 

\author{H. S. Vieira\inst{1} \and V. B. Bezerra\inst{1} \and Andr\'{e} A. Costa\inst{2}}
\shortauthor{H. S. Vieira \etal}

\institute{
	\inst{1} Departamento de F\'{i}sica, Universidade Federal da Para\'{i}ba, Caixa Postal 5008, CEP 58051-970, Jo\~{a}o Pessoa, PB, Brazil\\
	\inst{2} Instituto de F\'{i}sica, Universidade de S\~{a}o Paulo, Caixa Postal 66318, CEP 05315-970, S\~{a}o Paulo, SP, Brazil
}


\abstract{
We analyze the influence of the gravitational field produced by a slowly rotating black hole with a cosmic string along the axis of symmetry on a massive scalar field. Exact solutions of both angular and radial parts of the Klein-Gordon equation in this spacetime are obtained, and are given in terms of the confluent Heun functions. We emphasize the role of the presence of the cosmic string in these solutions. We also investigate the solutions in regions near and far from the event horizon. From the radial solution, we obtain the exact wave solutions near the exterior horizon of the black hole, and discuss the Hawking radiation of massive scalar particles.
}

\begin{document}

\maketitle

%
%
\section{Introduction}
The knowledge of the behavior of different fields which interact with the gravitational field of black holes, certainly, will give us some relevant informations about the physics of these objects. In particular the scalar field constitues one of these fields which may give us important informations about the physics of a black hole. Along this line of research, the separability of the Klein-Gordon equation has been studied in different black hole backgrounds \cite{CommMathPhys.10.280,PhysRevD.5.1913,ClassQuantumGrav.22.339}. Others studies include the determination of solutions of the Klein-Gordon equation in different black hole gravitational fields as well as their consequences \cite{PhysRevD.12.2963,JMathPhys.22.1457,semiz,ClassQuantumGrav.31.045003}.

The cosmic string is a linear topological defect predicted in some gauge field theories \cite{JPhysAMathGen.9.1387} as a result of phase transitions and are topologically stable. It can either form closed loops or extend to infinity, and is characterized by its tension, $G\mu$, where $G$ is the Newton's gravitational constant and $\mu$ is the mass per unit length of the string \cite{Vilenkin:1994}.

The spacetime geometry associated with a static, straight infinite and infinitely thin cosmic string has a conical structure. The spacetime metric is given by $ds^{2}=dt^{2}-dr^{2}-(1-4G\mu)^{2}r^{2}\ d\phi^{2}-dz^{2}$. This metric can be obtained from the one corresponding to a string with a given radius by letting this radius go to zero and keeping the parameter $\mu$ constant.

It is locally flat, except on the localization of the string where it has a delta shaped curvature tensor. The section perpendicular to the cosmic string has an azimuthal deficit angle given by $\Delta \phi=8 \pi G \mu$. This means that this spacetime is locally flat but not globally. The local flatness of the spacetime surrounding a cosmic string means that there is no local gravitational force. However, there exist some interesting gravitational effects associated with the non-trivial topology of the spacelike section arround the cosmic string. Among these effects, a cosmic string can act as a gravitational lens, it can induce a finite electrostatic self-force on an eletric charged particle, the emission of radiation by a freely moving particle, and many other effects \cite{PhysRevD.23.852}.

In principle, a cosmic string can appear not as a single object in empty space, but rather as part of a larger gravitational system, as for example, passing through a black hole. In general, solutions of Einstein's equations can be constructed by including a cosmic string \cite{ClassQuantumGrav.6.1313}, where some investigations were done revealing the role played by the cosmic string. In this case, the spacetime with a cosmic string can be constructed by removing a wedge, that is, by requiring that the azimuthal angle around the axis of symmetry runs over the range $0 < \phi < 2 \pi b$, with $b=1-4G\mu$. Then, gluing together the resulting edges we get the spacetime with a cosmic string.

Hawking radiation is an interesting phenomenon which corresponds to a spontaneous emission of black body radiation by black holes, whose original prediction was done by calculating the thermal radiation emitted by a spherically symmetric black hole \cite{CommunMathPhys.43.199}. The studies concerning this phenomenon were carried on by using different methods \cite{PhysRevLett.85.5042,EPL.100.49001,EPL.107.50001}. In particular, the emission of scalar particles by black holes has been an object of investigation in recent years \cite{AnnPhys.350.14}.

In this letter, we obtain the exact solutions of the Klein-Gordon equation for a massive scalar field in the Lense-Thirring spacetime with a cosmic string passing through it, valid in the whole space that corresponds to the region between the exterior event horizon and infinity. They are given in terms of solutions of the Heun equations \cite{Ronveaux:1995}. In this sense, we extend the range in which the solutions are valid as compared with the ones obtained in the literature \cite{bezerra}. We also analyze the asymptotic behavior of the solutions. Using the radial solution which is given in terms of the confluent Heun functions and taking into account their properties, we study the Hawking radiation of massive scalar particles.
%
%
\section{Lense-Thirring spacetime with a cosmic string}
The metric generated by a black hole with angular momentum per mass, $a$, and mass (energy), $M$, in the limit of slow rotation, that is, $a^2 \approx 0$, which corresponds to the Lense-Thirring metric \cite{PhysZ.19.156}, whose line element, in the Boyer-Lindquist coordinates, is given by
\begin{eqnarray}
ds^{2} & = & \frac{\Delta}{r^{2}}\ dt^{2}-\frac{r^{2}}{\Delta}\ dr^{2}-r^{2}\ d\theta^{2}-r^{2}\sin^{2}\theta\ d\phi^{2}\nonumber\\
& + & \frac{4 a M \sin^{2}\theta}{r}\ dt\ d\phi\ ,
\label{eq:metrica_Lense-Thirring}
\end{eqnarray}
where $\Delta=r^{2}-2Mr$. The line element above is the first approximation of the Kerr solution, valid up to first order in $a/r$. The order of approximation considered simplifies our result and at the same time is enough for our proposals, because what we want is to show up the effect of rotation combined with the one produced by the cosmic string.

Now, let us introduce a cosmic string in the Lense-Thirring spacetime. This can be made simply redefining the azimuthal angle $\phi$ in such a way that $\phi \rightarrow b\phi$. Doing this, we obtain the metric corresponding to a Lense-Thirring spacetime with a cosmic string passing through it, which can be rewritten as \cite{ClassQuantumGrav.6.1313}
\begin{eqnarray}
ds^{2} & = & \frac{\Delta}{r^{2}}\ dt^{2}-\frac{r^{2}}{\Delta}\ dr^{2}-r^{2}\ d\theta^{2}-b^{2}r^{2}\sin^{2}\theta\ d\phi^{2}\nonumber\\
& + & \frac{4 a M b\sin^{2}\theta}{r}\ dt\ d\phi\ ,
\label{eq:metrica_Lense-Thirring_string}
\end{eqnarray}
where the parameter $b$, which codifies the presence of the cosmic string, is such that $0 < b < 1$.

In the particle models which predict cosmic string solutions, these topological defects appear, in fact, as a set of infinite strings as well as string loops. In this paper we will consider just a static, infinite and straight cosmic string and as consequence we will not take into account the phenomena concerning gravitational accretion nor the formation of wiggly strings. Therefore, we can consider the total mass of the system under consideration as a constant.

From Eq.~(\ref{eq:metrica_Lense-Thirring_string}), we have that the horizon surface equation is obtained from the condition
\begin{equation}
\Delta=(r-r_{+})(r-r_{-})=0\ .
\label{eq:superficie_hor_Lense-Thirring}
\end{equation}
The solutions of Eq.~(\ref{eq:superficie_hor_Lense-Thirring}) are
\begin{equation}
r_{+}=2M,\ r_{-}=0\ ,
\label{eq:sol_padrao_Lense-Thirring_1}
\end{equation}
and correspond to the event and Cauchy horizons of the background under consideration.

The gravitational acceleration on the background horizon surface, $r_{+}$, is
\begin{equation}
\kappa_{+} \equiv \frac{1}{2r_{+}^{2}}\left.\frac{d\Delta}{dr}\right|_{r=r_{+}}=\frac{1}{2r_{+}}=\frac{1}{4M}\ .
\label{eq:acel_grav_ext_Lense-Thirring}
\end{equation}
%
%
\subsection{Thermodynamics}
Let us derive the thermodynamic quantities associated with the Lense-Thirring black hole pierced by a cosmic string along the axis of rotation.

The Hawking radiation temperature in geometric units, $T_{+}$, and the surface area of the horizon in the presence of a cosmic string, $\mathcal{A}_{+,b}$, are given by
\begin{equation}
T_{+}=\frac{\kappa_{+}}{2\pi}=\frac{1}{8 \pi M}\ ,
\label{eq:temp_Hawking_Lense-Thirring}
\end{equation}
\begin{equation}
\mathcal{A}_{+,b}=\left.\int\int\sqrt{-g}\ d\theta\ d\phi \right|_{r=r_{+}}=4 \pi r_{+}^{2}b\ ,
\label{eq:area_Lense-Thirring}
\end{equation}
where $g \equiv \mbox{det}(g_{\sigma\tau})=-b^{2}r^{4}\sin^{2}\theta$. Hence, the entropy at the event horizon in the presence of the cosmic string, $S_{+,b}$, is given by
\begin{equation}
S_{+,b}=\frac{\mathcal{A}_{+,b}}{4}=\pi r_{+}^{2}b=4 \pi M^{2}b\ .
\label{eq:entropia_Lense-Thirring}
\end{equation}
From Eq.~(\ref{eq:metrica_Lense-Thirring_string}), the dragging angular velocity of the exterior horizon in the presence of the cosmic string, $\Omega_{+,b}$, is given by
\begin{equation}
\Omega_{+,b}=\left.-\frac{g_{03}}{g_{33}}\right|_{r=r_{+}}=\frac{a}{r_{+}^{2}b}=\frac{a}{4M^{2}b}\ .
\label{eq:vel_ang_Lense-Thirring}
\end{equation}
The Komar's mass and angular momentum are
\begin{equation}
M_{phys}=Mb,\ J_{phys}=Jb\ ,
\label{eq:massa_Lense-Thirring}
\end{equation}
so that the ratio $a=J/M=J_{phys}/M_{phys}$ remains unchanged. From Eq.~(\ref{eq:massa_Lense-Thirring}) we obtain
\begin{equation}
Eb=E_{phys}\ ,
\label{eq:energia_Lense-Thirring}
\end{equation}
which means that the physical energy of the Lense-Thirring spacetime is decreased due to the presence of a cosmic string.

These quantities given by Eqs.~(\ref{eq:temp_Hawking_Lense-Thirring})-(\ref{eq:energia_Lense-Thirring}) for the event horizon of the black hole with a cosmic string satisfy the first law of thermodynamics, namely,
\begin{equation}
dE_{phys}=T_{+}\ dS_{+,b}+\Omega_{+,b}\ dJ_{phys}\ .
\label{eq:1_lei_termo_Lense-Thirring}
\end{equation}
%
%
\section{The Klein-Gordon equation in a Lense-Thirring spacetime with a cosmic string}
We want to study the behavior of scalar fields in the curved spacetime generated by a slowly rotating black hole with a cosmic string passing through it, along the axis of symmetry. To do this, we must solve the covariant Klein-Gordon equation, which has the form
\begin{equation}
\left[\frac{1}{\sqrt{-g}}\partial_{\sigma}(g^{\sigma\tau}\sqrt{-g}\partial_{\tau})+\mu_{0}^{2}\right]\Psi=0\ ,
\label{eq:Klein-Gordon_cova}
\end{equation}
where $\mu_{0}$ is the mass and the units $c \equiv \hbar \equiv 1$ were chosen. From now on we will also choose $G=1$.

Due to the time independence and symmetry of the spacetime with respect to rotation, the solution of Eq.~(\ref{eq:Klein-Gordon_cova}) can be written as
\begin{equation}
\Psi=\Psi(\mathbf{r},t)=R(r)S(\theta)\mbox{e}^{im\phi}\mbox{e}^{-i \omega t}\ ,
\label{eq:separacao_variaveis}
\end{equation}
where $m=\pm 1,\pm 2, \pm 3,...$ is the azimuthal quantum number. Substituting Eqs.~(\ref{eq:metrica_Lense-Thirring_string}) and (\ref{eq:separacao_variaveis}) into Eq.~(\ref{eq:Klein-Gordon_cova}), this equation can be separated into two equations for $S(\theta)$ and $R(r)$, namely,
\begin{equation}
\frac{1}{\sin\theta}\frac{d}{d\theta}\left(\sin\theta\frac{dS}{d\theta}\right)+\left(\lambda-\frac{m^{2}}{b^{2}\sin^{2}\theta}\right)S=0\ ,
\label{eq:mov_angular_Lense-Thirring}
\end{equation}
\begin{equation}
\frac{d}{dr}\left(\Delta\frac{dR}{dr}\right)+\biggl[\frac{\omega^{2}r^{4}}{\Delta}-\frac{4Ma\omega mr}{b\Delta}-(\mu_{0}^{2}r^{2}+\lambda)\biggr]R=0\ ,
\label{eq:mov_radial_1}
\end{equation}
where $\lambda$ is a constant. In what follows we will solve exactly the angular and radial parts of the Klein-Gordon equation, given by Eqs.~(\ref{eq:mov_angular_Lense-Thirring}) and (\ref{eq:mov_radial_1}), respectively.
%
%
\subsection{Angular equation}
Note that the presence of the cosmic string modifies slightly the angular part of the Klein-Gordon equation with respect to the case without the cosmic string which can be obtained by simply taking $b=1$. This modification is reflected in the constant $m$, which turns to be $m_{(b)} \equiv m/b$. In the specific case in which $b=1$, Eq.~(\ref{eq:mov_angular_Lense-Thirring}) has solutions which corresponds to the associated Legendre polynomials, $P_{l}^{m}(\cos\theta)$, with eigenvalues $\lambda_{lm}=l(l+1)$, where $l,m$ are integers such that $|m| \leq l$. It can be shown that the function $S=S(\theta)$ which solves Eq.~(\ref{eq:mov_angular_Lense-Thirring}) for $b \neq 1$, has solutions that can be expressed in terms of the confluent Heun functions.

We can rewrite Eq.~(\ref{eq:mov_angular_Lense-Thirring}) in the form which resembles a Heun equation, by defining a new angular coordinate $x=\cos^{2}\theta$, as
\begin{eqnarray}
\frac{d^{2}S}{dx^{2}}+\left(\frac{1/2}{x}+\frac{1}{x-1}\right)\frac{dS}{dx}+\biggl\{\frac{\lambda-m_{(b)}^{2}}{4}\frac{1}{x}\nonumber\\
+\frac{m_{(b)}^{2}-\lambda}{4}\frac{1}{x-1}-\left[\frac{m_{(b)}}{2}\right]^{2}\frac{1}{(x-1)^{2}}\biggr\}S=0\ .
\label{eq:mov_angular_x_Lense-Thirring_3}
\end{eqnarray}

Now, let us perform a transformation in order to reduce the power of the term proportional to $1/(x-1)^{2}$. This is a \textit{s-homotopic transformation} of the dependent variable, $S(x)=(x-1)^{A_{1}}U(x)$, where $A_{1}=m_{(b)}/2$. The function $U(x)$ satisfies the following equation
$$
\frac{d^{2}U}{dx^{2}}+\left(\frac{1/2}{x}+\frac{2 A_{1} +1}{x-1}\right)\frac{dU}{dx}\nonumber\\
$$
\begin{equation}
+\biggl\{\frac{[-2 A_{1}+\lambda -m_{(b)}^{2}]/4}{x}+\frac{[2 A_{1}-\lambda +m_{(b)}^{2}]/4}{x-1}\biggr\}U=0\ ,
\label{eq:mov_angular_x_Lense-Thirring_4}
\end{equation}
which is similar to the confluent Heun equation \cite{JPhysAMathTheor.43.035203} given by
\begin{equation}
\frac{d^{2}U}{dx^{2}}+\left(\alpha+\frac{\beta+1}{x}+\frac{\gamma+1}{x-1}\right)\frac{dU}{dx}+\left(\frac{\mu}{x}+\frac{\nu}{x-1}\right)U=0\ ,
\label{eq:Heun_confluente_forma_canonica}
\end{equation}
where $U(x)=\mbox{HeunC}(\alpha,\beta,\gamma,\delta,\eta;x)$ are the confluent Heun functions, with the parameters $\alpha$, $\beta$, $\gamma$, $\delta$ and $\eta$, related to $\mu$ and $\nu$ by $\mu=(\alpha-\beta-\gamma+\alpha\beta-\beta\gamma)/2-\eta$ and $\nu=(\alpha+\beta+\gamma+\alpha\gamma+\beta\gamma)/2+\delta+\eta$.

Thus, the general solution of the angular part of the Klein-Gordon equation for a massive scalar field in the Lense-Thirring spacetime with a cosmic string, in the exterior region of the event horizon, given by Eq.~(\ref{eq:mov_angular_x_Lense-Thirring_3}) over the entire range $0 \leq x < \infty$, can be written as
\begin{eqnarray}
S(x) & = & (x-1)^{\frac{1}{2}\gamma}\nonumber\\
& \times & \{C_{1}\ \mbox{HeunC}(\alpha,\beta,\gamma,\delta,\eta;x)\nonumber\\
& + & C_{2}\ x^{-\beta}\ \mbox{HeunC}(\alpha,-\beta,\gamma,\delta,\eta;x)\}\ ,
\label{eq:solucao_geral_radial_Schwarzschild}
\end{eqnarray}
where $C_{1}$ and $C_{2}$ are constants, and the parameters $\alpha$, $\beta$, $\gamma$, $\delta$, and $\eta$ are now given by:
\begin{equation}
\alpha=0;\ \beta=-\frac{1}{2};\ \gamma=m_{(b)};\ \delta=0;\ \eta=\frac{1+m_{(b)}^{2}-\lambda}{4}\ .
\label{eq:alpha-beta_radial_HeunC_Schwarzschild}
\end{equation}
These two functions form linearly independent solutions of the confluent Heun dif\-fer\-en\-tial equation, provided $\beta$ is not integer. Note the dependence of the angular solution with the parameter $b$, associated with the presence of the cosmic string.
%
%
\subsection{Radial equation}
In order to solve the radial part of the Klein-Gordon equation, let us use Eq.~(\ref{eq:superficie_hor_Lense-Thirring}) and define a new coordinate, $x$, such that \cite{bezerra}
\begin{equation}
Mx=r-r_{+}=r-2M\ .
\label{eq:x_Lense-Thirring}
\end{equation}
The new coordinate is such that when $r \rightarrow r_{+}$, $x \rightarrow 0$, and when $r \rightarrow \infty$, $x \rightarrow \infty$. We are interested to study the behavior of the fields only in the exterior region to the event horizon, namely, for $r > r_{+}$, which corresponds to $0 \leq x < \infty$, in terms of the new coordinate.

Now, defining a new function by $R(x)=Z(x)[x(x+2)]^{-1/2}$, and changing the variable from $x$ to $z$, through the relation
\begin{equation}
z=-\frac{x}{2}\ ,
\label{eq:variavel_z_Lense-Thirring}
\end{equation}
Eq.~(\ref{eq:mov_radial_1}) can be written as
\begin{eqnarray}
\frac{d^{2}Z}{dz^{2}}+\biggl[4M^{2}(\omega^{2}-\mu_{0}^{2})+\frac{A}{M^{2}z^{2}}+\frac{-2B}{M^{2}z}\nonumber\\
+\frac{C}{M^{2}(z-1)^{2}}+\frac{-2D}{M^{2}(z-1)}\biggr]Z=0\ ,
\label{eq:radial_Lense-Thirring_normal}
\end{eqnarray}
where the coefficients $A$, $B$, $C$, and $D$ are given by:
\begin{eqnarray}
&&A=\frac{1}{4}\left[M^{2}-8m_{(b)}a\omega M^{2}+16\omega^{2}M^{4}\right];\nonumber\\
&&B=\frac{1}{4}[-M^{2}-2M^{2}\lambda-8\mu_{0}^{2}M^{4}+4am_{(b)}\omega M^{2}\nonumber\\
&&+16\omega^{2}M^{4}];\nonumber\\
&&C=\frac{M^{2}}{4};\nonumber\\
&&D=\frac{1}{4}[M^{2}+2M^{2}\lambda-4am_{(b)}\omega M^{2}]\ .
\label{eq:A-D}
\end{eqnarray}

The normal form of the confluent Heun equation is given by \cite{Ronveaux:1995}
\begin{equation}
\frac{d^{2}Z}{dx^{2}}+\left[B_{1}+\frac{B_{2}}{x^{2}}+\frac{B_{3}}{x}+\frac{B_{4}}{(x-1)^{2}}+\frac{B_{5}}{x-1}\right]Z=0\ ,
\label{eq:Heun_confluente_forma_normal}
\end{equation}
with
\begin{equation}
Z(x)=U(x)\mbox{e}^{\frac{1}{2}\int\left(\alpha+\frac{\beta+1}{x}+\frac{\gamma+1}{x-1}\right)dx}\ ,
\label{eq:solZ}
\end{equation}
where the coefficients $B_{1}$, $B_{2}$, $B_{3}$, $B_{4}$, and $B_{5}$ are given by:
\begin{eqnarray}
&&B_{1} \equiv -\frac{1}{4}\alpha^{2};\ B_{2} \equiv \frac{1}{4}(1-\beta^{2});\ B_{3} \equiv\frac{1}{2}(1-2\eta);\nonumber\\
&&B_{4} \equiv \frac{1}{4}(1-\gamma^{2});\ B_{5} \equiv \frac{1}{2}(-1+2\delta+2\eta)\ .
\label{eq:B1-B5_Heun_confluente_forma_normal}
\end{eqnarray}

Thus, the general solution of the radial part of the Klein-Gordon equation for a massive scalar field in the Lense-Thirring spacetime with a cosmic string, in the exterior region of the event horizon, given by Eq.~(\ref{eq:radial_Lense-Thirring_normal}) over the entire range $0 \leq x < \infty$, can be written as
\begin{eqnarray}
R(z) & = & \frac{M}{\Delta^{1/2}}\mbox{e}^{\frac{1}{2}\alpha z}(z-1)^{\frac{1}{2}(1+\gamma)}z^{\frac{1}{2}(1+\beta)}\nonumber\\
& \times & \{C_{1}\ \mbox{HeunC}(\alpha,\beta,\gamma,\delta,\eta;z)\nonumber\\
& + & C_{2}\ z^{-\beta}\ \mbox{HeunC}(\alpha,-\beta,\gamma,\delta,\eta;z)\}\ ,
\label{eq:solucao_geral_radial_Lense-Thirring}
\end{eqnarray}
where $C_{1}$ and $C_{2}$ are constants, and the parameters $\alpha$, $\beta$, $\gamma$, $\delta$, and $\eta$ are now given by:
\begin{eqnarray}
&&\alpha=4M(\mu_{0}^{2}-\omega^{2})^{1/2};\ \beta=i4M\left[\omega-\frac{am_{(b)}}{4M^{2}}\right];\ \gamma=0;\nonumber\\
&&\delta=4M^{2}\left(\mu_{0}^{2}-2\omega^{2}\right)\ ;\nonumber\\
&&\eta=-4M^{2}\left(\mu_{0}^{2}-2\omega^{2}\right)+2am_{(b)}\omega-\lambda\ .
\label{eq:alpha_HeunC_Lense-Thirring}
\end{eqnarray}

These two functions form linearly independent solutions of the confluent Heun differential equation provided $\beta$ is not integer, which is consistent with the fact that there is not any specific physical reason to impose that $\beta$ should be an integer.

Now, let us analyze the asymptotic behavior of the general solution of Eq.(\ref{eq:radial_Lense-Thirring_normal}), given by Eq.(\ref{eq:solucao_geral_radial_Lense-Thirring}), over the range $0 \leq x < \infty$. Firstly, we will consider a region close to the event horizon, which means that $r \rightarrow r_{+}$, that is, $x \rightarrow 0$. Secondly, we will consider a region very far from the black hole, that is, $r \rightarrow \infty$, or equivalently, $x \rightarrow \infty$.

It is important to take these limits into account in order to obtain the appropriate solutions to study the black hole radiation, in which case we need to know the outgoing wave at the horizon surface $r=r_{+}$.
%
%
\subsection{Case 1}
When $z \rightarrow 0$, we have that $x \rightarrow 0$ and $r \rightarrow r_{+}$. Thus, using the expansion in power series of the confluent Heun functions with respect to the independent variable $z$, in a neighborhood of the regular singular point $z=0$ \cite{Ronveaux:1995}, we get
\begin{equation}
\mbox{HeunC}(\alpha,\beta,\gamma,\delta,\eta;z)=1+\frac{1}{2}a_{1}z+...\ ,
\label{eq:serie_HeunC_todo_z}
\end{equation}
where $a_{1}=(-\alpha\beta+\beta\gamma+2\eta-\alpha+\beta+\gamma)/(\beta+1)$. The solutions of (\ref{eq:radial_Lense-Thirring_normal}), in this limit, are given by
\begin{equation}
Z(x) \sim C_{1}\left(-\frac{x}{2}\right)^{\frac{1}{2}+\bar{m}}+C_{2}\left(-\frac{x}{2}\right)^{\frac{1}{2}-\bar{m}}\ ,
\label{eq:solucao_radial_caso1_limite}
\end{equation}
with $\bar{m}^{2}=1/4-A/M^{2}$.

At this point, we can compare the obtained result with the ones given in \cite{bezerra}. Note that there is a difference between the functional form of the asymptotic behavior of the general solution of the radial equation obtained analytically, given by Eq.(\ref{eq:solucao_radial_caso1_limite}), and the approximated solution obtained in \cite{bezerra}, which is given by
\begin{equation}
Z(x) \sim C_{1}\ \mbox{e}^{-\sqrt{F}x}(2\sqrt{F}x)^{\frac{1}{2}+\bar{m}}+C_{2}\ \mbox{e}^{-\sqrt{F}x}(2\sqrt{F}x)^{\frac{1}{2}-\bar{m}}\ ,
\label{eq:solucao_radial_caso1}
\end{equation}
where $F=M^{2}(\mu_{0}^{2}-\omega^{2})-C/4M^{2}-D/2M^{2}$. It is worth noticing that expanding Eq.~(\ref{eq:solucao_radial_caso1}) up to first order in terms of $x=(r-r_{+})/M$, we conclude that the results given by Eqs.~(\ref{eq:solucao_radial_caso1_limite}) and (\ref{eq:solucao_radial_caso1}) are in accordance. In fact, the formal difference concerns to a multiplicative constant, which should be adjusted appropriately.
%
%
\subsection{Case 2}
When $|z| \rightarrow \infty$, we have that $|x| \rightarrow \infty$ and $r \rightarrow \infty$. Thus, using the fact that in the neighborhood of the irregular singular point at infinity, the two solutions of the confluent Heun equation exist, in general they can be expanded (in a sector) in the following asymptotic series \cite{Ronveaux:1995}
\begin{equation}
\mbox{HeunC}(\alpha,\beta,\gamma,\delta,\eta;z) \sim z^{-\frac{\beta+\gamma+2}{2}}(C_{1}\ z^{-\frac{\delta}{\alpha}}+C_{2}\ z^{\frac{\delta}{\alpha}}\mbox{e}^{-\alpha z})\ ,
\label{eq:assintotica_HeunC_z_grande}
\end{equation}
where we are keeping only the first term of this power-series asymptotics. Thus, the solutions of Eq.~(\ref{eq:radial_Lense-Thirring_normal}), in this limit, are given by
\begin{eqnarray}
&&Z(x) \sim C_{1}\ \mbox{e}^{-M[(\mu_{0}^{2}-\omega^{2})^{1/2}x]}\left(-\frac{x}{2}\right)^{\kappa}\nonumber\\
&&+C_{2}\ \mbox{e}^{+M[(\mu_{0}^{2}-\omega^{2})^{1/2}x]}\left(-\frac{x}{2}\right)^{-\kappa}\ ,
\label{eq:solucao_radial_caso2_limite}
\end{eqnarray}
where $\kappa=(B+D)/2M^{3}(\mu_{0}^{2}-\omega^{2})^{1/2}$.

Therefore, at infinity, we have asymptotic forms which are consistent with the fact that very far from the black hole, the Lense-Thirring spacetime with a cosmic string tends to Minkowski spacetime minus a wedge.

In this case there is also a difference between Eq.(\ref{eq:solucao_radial_caso2_limite}) and the one obtained in \cite{bezerra}, which is given by
\begin{equation}
Z(x) \sim C_{1}\ \mbox{e}^{-\sqrt{G}x}\left(2\sqrt{G}x\right)^{\kappa}+C_{2}\ \mbox{e}^{+\sqrt{G}x}\left(2\sqrt{G}x\right)^{-\kappa}\ ,
\label{eq:solucao_radial_caso2}
\end{equation}
where $G=M^{2}(\mu_{0}^{2}-\omega^{2})$, provided $\mu_{0}^{2}-\omega^{2} \neq 0$. Once more, the two results are equivalent, except for a multiplicative constant.

Thus, we can conclude that all solutions of the Klein-Gordon equation presented here, both angular and radial parts, depend on the parameter $b$ that defines the conicity of the Lense-Thirring spacetime due to the presence of the cosmic string. This means that the global aspects of this spacetime associated with the cosmic string are codified in these solutions.
%
%
\section{Hawking radiation and analytic extension}
From Eqs.~(\ref{eq:x_Lense-Thirring}), (\ref{eq:variavel_z_Lense-Thirring}) and (\ref{eq:serie_HeunC_todo_z}) we can see that the radial solution given by Eq.~(\ref{eq:solucao_geral_radial_Lense-Thirring}), near the exterior event horizon, that is, when $r \rightarrow r_{+}$, which implies that $x \rightarrow 0$, behaves asymptotically as
\begin{equation}
R(r) \sim C_{1}\ (r-r_{+})^{\beta/2}+C_{2}\ (r-r_{+})^{-\beta/2}\ ,
\label{eq:exp_0_solucao_geral_radial_Lense-Thirring}
\end{equation}
where we are considering contributions only of the first term in the expansion, and all constants are included in $C_{1}$ and $C_{2}$. Thus, considering the time factor, near the black hole event horizon $r_{+}$, this solution is given by
\begin{equation}
\Psi=\mbox{e}^{-i \omega t}(r-r_{+})^{\pm\beta/2}\ .
\label{eq:sol_onda_radial_Lense-Thirring}
\end{equation}
Substituting Eqs.~(\ref{eq:acel_grav_ext_Lense-Thirring}) and (\ref{eq:vel_ang_Lense-Thirring}) into Eq.~(\ref{eq:alpha_HeunC_Lense-Thirring}), for the parameter $\beta$, we get
\begin{equation}
\frac{\beta}{2}=\frac{i}{2\kappa_{+}}(\omega-\omega_{0,b})\ ,
\label{eq:expoente_rad_Hawking_Lense-Thirring}
\end{equation}
where $\omega_{0,b}=m\Omega_{+,b}$.

Therefore, on the black hole exterior horizon surface, the ingoing and outgoing wave solutions are
\begin{equation}
\Psi_{in}=\mbox{e}^{-i \omega t}(r-r_{+})^{-\frac{i}{2\kappa_{+}}(\omega-\omega_{0,b})}\ ,
\label{eq:sol_in_1_Lense-Thirring}
\end{equation}
\begin{equation}
\Psi_{out}(r>r_{+})=\mbox{e}^{-i \omega t}(r-r_{+})^{\frac{i}{2\kappa_{+}}(\omega-\omega_{0,b})}\ .
\label{eq:sol_out_2_Lense-Thirring}
\end{equation}
These solutions depend on the parameter $b$, in such a way that the total energy of the radiated particles is decreased due to presence of the cosmic string. It is worth calling attention to the fact that we are using the analytical solution of the radial part of the Klein-Gordon equation in the spacetime under consideration, differently from the calculations usually done in the literature, as for example in \cite{PhysLettB.618.14,ChinJPhys.47.618}.

Using the definitions of the tortoise and Eddington-Finkelstein coordinates, and taking into account that $a^{2} \approx 0$, we get:
\begin{eqnarray}
&&dr_{*}=\frac{r^{2}}{\Delta}dr;\ \ln(r-r_{+})=\frac{1}{r_{+}^{2}}\left.\frac{d\Delta}{dr}\right|_{r=r_{+}}r_{*}=2\kappa_{+}r_{*};\nonumber\\
&&\hat{r}=\frac{\omega-\omega_{0,b}}{\omega}r_{*};\ v=t+\hat{r}\ .
\label{eq:coord_Eddington-Finkelstein}
\end{eqnarray}
In those new coordinates, we obtain the following ingoing and outgoing wave solutions:
\begin{equation}
\Psi_{in}=\mbox{e}^{-i \omega v}\ ;
\label{eq:sol_in_1_Lense-Thirring_tortoise}
\end{equation}
\begin{equation}
\Psi_{out}(r>r_{+})=\mbox{e}^{-i \omega v}(r-r_{+})^{\frac{i}{\kappa_{+}}(\omega-\omega_{0,b})}\ .
\label{eq:sol_out_2_Lense-Thirring_tortoise}
\end{equation}
The solutions (\ref{eq:sol_in_1_Lense-Thirring_tortoise}) and (\ref{eq:sol_out_2_Lense-Thirring_tortoise}), in the case $b=1$, are exactly the solutions obtained in our recent paper \cite{AnnPhys.350.14} (putting $Q=0$, and $a^{2} \approx 0$ in their Eqs.(96) and (97)).
%
%

From Eq.~(\ref{eq:sol_out_2_Lense-Thirring_tortoise}), we see that this solution is not analytical in the exterior event horizon $r=r_{+}$. By analytic continuation, rotating $-\pi$ through the lower-half complex $r$ plane, we obtain $(r-r_{+}) \rightarrow \left|r-r_{+}\right|\mbox{e}^{-i\pi}=(r_{+}-r)\mbox{e}^{-i\pi}$. Thus, the outgoing wave solution on the horizon surface $r_{+}$ is
\begin{equation}
\Psi_{out}(r<r_{+})=\mbox{e}^{-i\omega v}(r_{+}-r)^{\frac{i}{\kappa_{+}}(\omega-\omega_{0,b})}\mbox{e}^{\frac{\pi}{\kappa_{+}}(\omega-\omega_{0,b})}\ .
\label{eq:sol_1_out_4_Lense-Thirring}
\end{equation}
Equations (\ref{eq:sol_out_2_Lense-Thirring_tortoise}) and (\ref{eq:sol_1_out_4_Lense-Thirring}) describe the outgoing wave outside and inside of the black hole, respectively. Therefore, for an outgoing wave of a particle with energy $\omega$, and angular momentum $m$, the outgoing decay rate or the relative scattering probability of the scalar wave at the event horizon surface $r=r_{+}$ is given by
\begin{equation}
\Gamma_{+}=\left|\frac{\Psi_{out}(r>r_{+})}{\Psi_{out}(r<r_{+})}\right|^{2}=\mbox{e}^{-\frac{2\pi}{\kappa_{+}}(\omega-\omega_{0,b})}\ .
\label{eq:taxa_refl_Lense-Thirring}
\end{equation}
%
%
\section{Radiation spectrum}
After the black hole event horizon radiates particles with energy $\omega$ and angular momentum $m$, in order to consider the reaction of the radiation of the particle to the spacetime, we must replace $M_{phys},J_{phys}$ by $M_{phys}-\omega,J_{phys}-m$, respectively, in the line element of the spacetime under consideration. Doing these changes, we must guarantee that the total energy and angular momentum of the spacetime are both conserved, that is,
\begin{equation}
-\omega=\Delta E_{phys},\ -m=\Delta J_{phys}\ ,
\label{eq:param_cons_Lense-Thirring}
\end{equation}
where $\Delta E_{phys}$ and $\Delta J_{phys}$ are the physical energy and Komar's angular momentum variations of the black hole event horizon, before and after the emission of radiation, respectively. Substituting Eqs.~(\ref{eq:1_lei_termo_Lense-Thirring}) and (\ref{eq:param_cons_Lense-Thirring}) into Eq.(\ref{eq:taxa_refl_Lense-Thirring}), we obtain the outgoing decay rate at the event horizon surface $r=r_{+}$:
\begin{equation}
\Gamma_{+}=\mbox{e}^{-\frac{2\pi}{\kappa_{+}}(-\Delta E_{phys}-m\Omega_{+,b})}=\mbox{e}^{\Delta S_{+,b}}\ ,
\label{eq:taxa_refl_param_Lense-Thirring}
\end{equation}
where $\Delta S_{+,b}$ is the change of the Bekenstein-Hawking entropy in the presence of the cosmic string, if we compare situations before and after the emission of radiation, being given by
\begin{equation}
\Delta S_{+,b}=-4 \pi \omega(2M-\omega)b\ .
\label{eq:entropia_Bekenstein-Hawking_Lense-Thirring}
\end{equation}
Now, let us consider the Damour-Ruffini-Sannan method \cite{PhysRevD.14.332,GenRelativGravit.20.239} which is valid to calculate the Hawking radiation for static as well for stationary black holes and therefore for a spacetime which is not globally hyperbolic. Thus, we get the following Hawking radiation spectrum of scalar particles
\begin{equation}
\left|N_{\omega}\right|^{2}=\frac{1}{\mbox{e}^{\frac{2\pi}{\kappa_{+}}(\omega-\omega_{0,b})}-1}=\frac{1}{\mbox{e}^{\frac{\hbar(\omega-\omega_{0,b})}{k_{B}T_{+}}}-1}\ .
\label{eq:espectro_rad_Lense-Thirring_2}
\end{equation}

Therefore, we can see that the resulting Hawking radiation spectrum of scalar particles has a thermal character, analogous to the black body spectrum. It is worth noticing that the total energy of radiated scalar particles is decreased due to the presence of the cosmic string, more precisely, the dragging angular velocity of the exterior horizon, $\Omega_{+,b}$, is amplified in comparison with the scenario without a cosmic string \cite{AnnPhys.350.14}.
%
%
\section{Conclusions}
These solutions extend the ones obtained in \cite{bezerra}, in the sense that now we have analytic solutions for all spacetime, which means, in the region between the event horizon and infinity, differently from the results obtained in \cite{bezerra} which are valid only in asymptotic regions, namely, very close to the horizons and far from the black hole. The radial solution is given in terms of the confluent Heun functions, and is valid over the range $0 \leq z < \infty$. As to the solution of the angular part, it also is given in terms of the confluent Heun functions.

We obtained the solutions for ingoing and outgoing waves near the exterior horizon of a Lense-Thirring black hole with a cosmic string, and used these results to discuss the Hawking radiation, in which we considered the properties of the confluent Heun functions to obtain the results. This approach has the advantage that it is not necessary to introduce any coordinate system, as for example, the particular one \cite{PhysLettB.618.14}, the tortoise or Eddington-Finkelstein coordinates \cite{ChinJPhys.47.618}. As the dragging angular velocity of the exterior horizon, $\Omega_{+,b}$, depends on the conicity, this quantity codifies the presence of the cosmic string, and in fact, is amplified by the presence of this topological defect.

From the local point of view, the gravitational field associated with the Lense-Thirring black hole with a cosmic string remains axially symmetric. But globally this symmetry was broken, and this fact produces some modifications in the physical states of the particles. The wave function depends on the parameter $b$ that codifies the presence of the string, and as a consequence all other physical quantities are also influenced by the presence of the cosmic string.
%
%
\acknowledgments
The authors would like to thank Conselho Nacional de Desenvolvimento Cient\'{i}fico e Tecnol\'{o}gico (CNPq) for partial financial support.
%
%

%
%
\end{document}